\documentclass[preprint,review,12pt]{elsarticle}

\usepackage{amssymb}
\usepackage{amsmath}
\usepackage[ruled,vlined,linesnumbered,titlenotnumbered]{algorithm2e}
\usepackage{xcolor}
\definecolor{arrowgreen}{RGB}{130,178,103}
\usepackage{booktabs}
\usepackage{color}
\usepackage{hyperref}
\usepackage{tablefootnote}

\usepackage{etoolbox,environ}

\newtoggle{updatecolor}
\NewEnviron{update}
  {\iftoggle{updatecolor}{\textcolor{black}{\BODY \space}}{\BODY}}
  
\toggletrue{updatecolor} 

\journal{Applications in Energy and Combustion Science}

\begin{document}

\begin{frontmatter}

\title{Deep Neural Networks to Correct Sub-Precision Errors in CFD}

\author[inst1]{Akash Haridas}
\affiliation[inst1]{organization={Indian Institute of Technology Madras},
            addressline={Dept. of Aerospace Engineering}, 
            postcode={600036}, 
            state={Chennai},
            country={India}}
\author[inst1]{Nagabhushana Rao Vadlamani\corref{cor1}}

\cortext[cor1]{Corresponding author: nrv@iitm.ac.in}

\author[inst2,inst3]{Yuki Minamoto}
\affiliation[inst2]{organization={Tokyo Institute of Technology},
            addressline={2-12-1 O-okayama}, 
            city={Meguro},
            postcode={152-8550}, 
            state={Tokyo},
            country={Japan}}
\affiliation[inst3]{organization={Japan Science and Technology Agency, PRESTO},
            addressline={7},
            city={Chiyoda},
            postcode={102-0076}, 
            state={Tokyo},
            country={Japan}}

\begin{abstract}
Information loss in numerical physics simulations can arise from various sources when solving discretized partial differential equations. In particular, errors related to numerical precision (\emph{"sub-precision errors"}) can accumulate in the quantities of interest when the simulations are performed using low-precision 16-bit floating-point arithmetic compared to an equivalent 64-bit simulation. On the other hand, low-precision computation is less resource intensive than high-precision computation. Several machine learning techniques proposed recently have been successful in correcting errors due to coarse spatial discretization. In this work, we extend these techniques to improve CFD simulations performed with low numerical precision. We quantify the precision-related errors accumulated in a Kolmogorov forced turbulence test case. Subsequently, we employ a Convolutional Neural Network together with a fully differentiable numerical solver performing 16-bit arithmetic to learn a tightly-coupled ML-CFD hybrid solver\protect\footnotemark. Compared to the 16-bit solver, we demonstrate the efficacy of the hybrid solver towards improving various metrics pertaining to the statistical and pointwise accuracy of the simulation.
\end{abstract}

\begin{graphicalabstract}
\includegraphics[width=\textwidth]{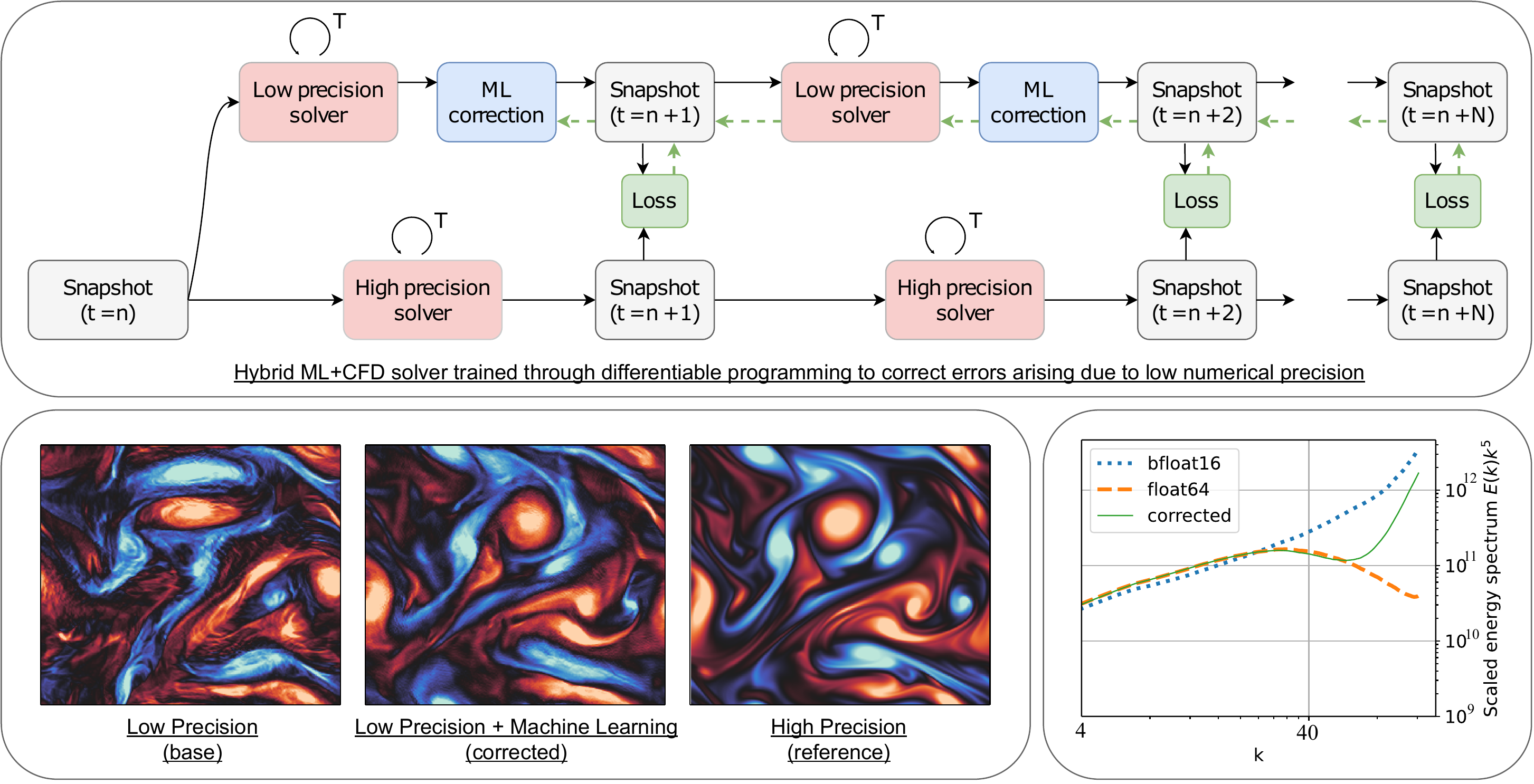}
\end{graphicalabstract}

\begin{highlights}
\item Developed a machine learning (ML) model to correct the low precision errors in CFD.
\item ML model couples a neural network with a fully differentiable CFD solver.
\item Achieved accurate solutions of the velocity field with improved visual fidelity.
\item Explored the effect of hyperparameters on computational cost vs accuracy trade-off.
\end{highlights}


\end{frontmatter}

\footnotetext{The code and models developed in this work are available at \url{https://github.com/akasharidas/subprecision-cfd}}


\section{Introduction}
Computational simulations of fluids and other complex physical systems have critical applications in engineering and the physical sciences. Accurate and dependable numerical simulations are essential for  designing aircraft components and turbomachinery, predicting climate patterns, and analysing flow in blood vessels. The mathematical models that describe these physical systems often come in the form of Partial Differential Equations (PDEs) represented in a discretised form in computers and are solved using various numerical schemes. The complexity of real-world systems being simulated presents a trade-off between resolving the finest spatio-temporal details and keeping the computational cost to a feasible limit. Information loss in a numerical simulation can arise from various sources: the spatial and temporal discretisation of the PDEs, or the availability of finite precision in the representation of numerical values in computer memory. This work examines the benefits of low numerical precision in Computational Fluid Dynamics (CFD) and uses deep learning to correct the resulting errors.

The rise of machine learning (ML) as a powerful tool for information processing, together with the large amounts of data generated in CFD simulations, has led to the use of ML for understanding, modelling, optimising and controlling fluid flows \citep{MLinFMsurvey}. Purely data-driven approaches to spatio-temporal forecasting learn deep neural networks in a supervised manner without enforcing constraints on the physical properties of the output \citep{xingjian2015convolutional, latentspacephysics, tompson2017accelerating}. Such methods are computationally efficient but fail to enforce constraints such as conservation of mass and momentum and are hence suitable for accelerating simulations in computer animations and games where the exact physical correctness of the flow is not essential. Loss-based approaches try to guide the neural network predictions by including prior scientific knowledge in the loss function. Physics-guided or physics-informed neural networks encode physical relationships between variables and underlying physical laws such as PDEs in the loss function, and hence are biased towards producing physically accurate outputs \cite{pgnn, pinn, wang2020towards}. Machine learning has also been used in turbulence closure modelling \cite{turbulence1, turbulence2, turbulence3}. 

Recent efforts to use machine learning in scientific computing have involved differentiable programming \cite{hu2019difftaichi, toussaint2018differentiable, de2018end}. Software frameworks for deep learning (e.g. PyTorch  and JAX) implement reverse-mode automatic differentiation to compute gradients in neural networks \cite{paszke2017automatic, jax2018github}. This capability can be leveraged by re-implementing numerical solvers in these frameworks, allowing a neural network to be trained end-to-end along with the numerical solver by backpropagating gradients through the solver. Recently this method has been demonstrated to improve generalisation to new inputs and the adherence to physical laws \cite{um2020solver, kochkov2021machine}. Since the gradients can be readily calculated end-to-end, a solver-ML system trained in this manner can also be used for flow control \cite{holl2020learning}, inverse design and shape optimization \cite{chen2021numerical,kumar2021inverse}. Several numerical solvers written in automatic differentiation frameworks have been recently open-sourced \cite{kochkov2021machine, holl2020phiflow, jaxmd2020}.

\begin{figure}[ptb]
\centerline{\includegraphics[width=0.5\textwidth]{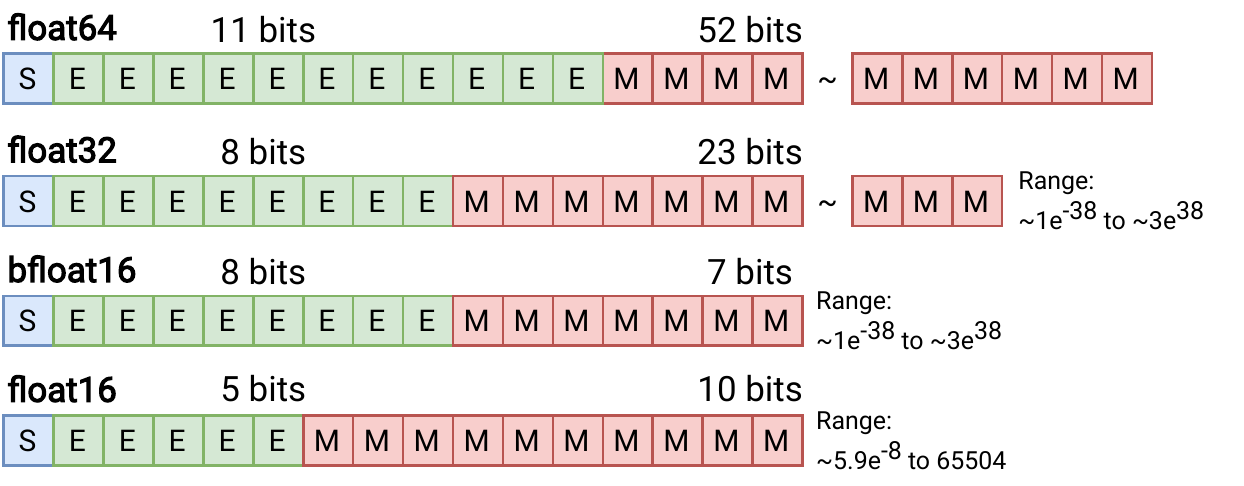}}
\caption{An illustration of the common floating-point formats used, including bfloat16. The number of exponent bits affects the available range, and the number of mantissa bits affects the precision. S, E, and M denote sign, exponent and mantissa bits, respectively.}
\label{fig:floating_point}
\end{figure}

Figure~\ref{fig:floating_point} illustrates the double-precision (float64) and single-precision (float32) floating-point formats commonly used in scientific computing. Most modern computer systems, including Graphics Processing Units (GPUs) often used for CFD, implement the IEEE 64-bit floating-point arithmetic standard, consisting of 53 mantissa bits resulting in roughly 15-17 significant digits of precision. The IEEE 32-bit floating-point standard consists of 23 mantissa bits resulting in approximately 6-9 significant digits of precision. In recent times, the realisation that deep neural networks retain their learning and predictive power even when using reduced-precision numerics has spurred the widespread adoption of floating-point standards such as half-precision IEEE 16-bit format (float16), which consists of 10 mantissa bits resulting in roughly 4 significant digits of precision \citep{mixedprecisiontraining}. This format is currently well supported on modern GPUs, with around 2-3 times more theoretical throughput than float32 \citep{abdelfattah2020survey}. The GPU memory savings obtained from the reduced numerical precision can also enable simulations with more mesh elements than would otherwise be possible. However, with 5 exponent bits, float16 has a maximum representable value of 65504. This can lead to overflow issues in scientific computing applications, including deep neural networks \citep{walden2019mixed}. The 'Brain' floating-point format (bfloat16) was developed at Google Brain to mitigate this issue \citep{kalamkar2019study}. The advantages of this format for computing are that the range of values it can represent is the same as that of float32, and conversion to and from float32 is simple, making it a drop-in replacement for float32 in situations where reduced precision is desired to trade-off precision and compute cost. The use of 16-bit formats is further motivated by a trend towards new hardware containing special units optimised for low-precision/mixed-precision computations. For example, bfloat16 is implemented in modern Intel x86 CPUs, Google's TPU accelerator and the new NVIDIA Ampere and Hopper architecture GPUs.

Previous studies on numerical precision in scientific computing noted that certain operations are more susceptible to error accumulation than others. In particular, large summations or accumulation operations, repeated operations in loops, large-scale parallel computation (since floating-point arithmetic is non-associative), and resolving small-scale phenomena cause the most error accumulation \citep{bailey2012high}. Simulations of turbulent fluid flows satisfy all of these conditions: resolving small-scale eddies is important; the PDEs are inherently non-linear, resulting in vastly different evolution trajectories with small changes in the initial conditions; and they are solved using large scale matrix operations on parallel processing devices such as GPUs. Climate modelling is an application where the accumulation of numerical error makes reproducibility and verification of the results difficult. \citet{he2001using} showed that employing double-double precision arithmetic in long inner product loops in atmospheric simulations dramatically reduces the variability of results between subsequent runs. 

To examine the benefits of reduced precision, we performed preliminary tests on the canonical test case of the Taylor Green Vortex on a cartesian grid of $2\pi^3$  \cite{debonis2013solutions}. We have used our in-house high-order CFD solver COMPSQUARE \cite{vadlamani2018distributed,achu2021entropically} to carry out these direct numerical simulations. The general purpose solver can handle body-fitted curvilinear meshes and uses high order compact schemes to solve the conservative form of the compressible Navier-Stokes equations. A V100 GPU with 32 GB of memory can accommodate a grid size of upto $415^3$ points ($\approx 72 \times 10^6$) using float32 and around $330^3$ points ($\approx 36 \times 10^6$) using float64. For a given grid, the simulations using float32 are $\approx 1.65 \times$ faster and can accommodate twice the grid size than those using float64. On hardware optimized for 16-bit computation, the speedup and the memory savings are expected to be even greater. For example, NVIDIA’s latest hardware optimized for 16-bit (and lower) precision is advertised to show between $3-12 \times$ speedup across all applications ranging from High Performance Computing (HPC) to Artificial Intelligence (AI) \cite{nvidia}.

This research is motivated by: (a) The speedup seen in low-precision simulations, (b) the reduced memory requirements, (c) the increasing hardware support for low-precision computation and (d) the demonstrated ability of deep learning to improve numerical simulations. 

In Section~\ref{sec:quantifying}, we quantify the error accumulation in a Kolmogorov forced turbulence test case. In Section~\ref{sec:reduce}, we explore deep learning strategies to correct this error. We present the key results from our experiments in Section~\ref{sec:results}, following which we conclude with a discussion on future directions in Section~\ref{sec:conclusion}.

\section{Quantifying the error accumulation}
\label{sec:quantifying}

\subsection{Test case and solver}

In this section, we first quantify the error accumulation due to reduced numerical precision and subsequently attempt to correct it using machine learning. For this, we simulate a 2D forced turbulence test case with JAX-CFD, an open-source CFD solver \citet{kochkov2021machine}. The solver uses a finite volume formulation for spatial discretisation on a staggered grid. Pressure is stored at the centre of each cell while the velocity components are defined on the corresponding faces. For diffusion, it uses a second-order central difference approximation of the Laplace operator. The advection term is solved using a second-order accurate scheme based on the Van Leer flux limiter \cite{vanleer}. The solver uses first-order Euler temporal discretisation and explicit time-stepping for advection and diffusion. Incompressible Navier-Stokes equations are solved using a pressure projection method. Depending on the dimensions of the grid, the code uses either a fast diagonalisation approach with explicit matrix multiplication \cite{pressureproj} or a real-valued fast Fourier transform. It is worth noting that the main scope of the present study is to develop, assess and demonstrate the framework of reducing the numerical error in low-precision CFD. Hence, the choice of relatively low-order schemes does not unduly affect the present conclusions. We use a $256 \times 256$ grid with periodic boundary conditions and set the Reynolds number (based on the domain dimension and initial maximum velocity) to 4000. In order to produce a statistically stationary turbulent flow, we apply the Kolmogorov forcing function \cite{kolomogorovflow} $\boldsymbol{f}$ given by the equation:

\[\boldsymbol{f} = (\sin(4y) - 0.1u) \boldsymbol{\hat{x}}\]

where $y$ is the vertical coordinate, $\boldsymbol{\hat{x}}$ is the unit vector along the horizontal axis and $u$ is the component of velocity along $\boldsymbol{\hat{x}}$. These forcing terms are added to the right-hand side of the $x$ and $y$ momentum Navier-Stokes equations. We note that in sub-grid modelling \cite{kochkov2021machine}, the error between the coarse and fine-grid simulations is considerable. On the other hand, in the present study, the forcing function ensures that the turbulent flow does not decay in time thereby resulting in a noticeable error accumulation with time.

The open-source solver is implemented in JAX \cite{jax2018github}, a high-performance numerical computing library. This allows us to run the simulation on GPUs and efficiently calculate gradients using automatic differentiation. We run this solver using either float64, float32 or bfloat16 numerics by setting the datatype of the initial velocity field. A run that uses float64 is treated as the reference high-fidelity simulation, and a run that starts at the same initial velocity field downcast to float32 or bfloat16 is treated as the corresponding low-fidelity simulation.

\subsection{Low-precision simulation characteristics}

\begin{figure*}[ptb]
\centerline{\includegraphics[width=\textwidth]{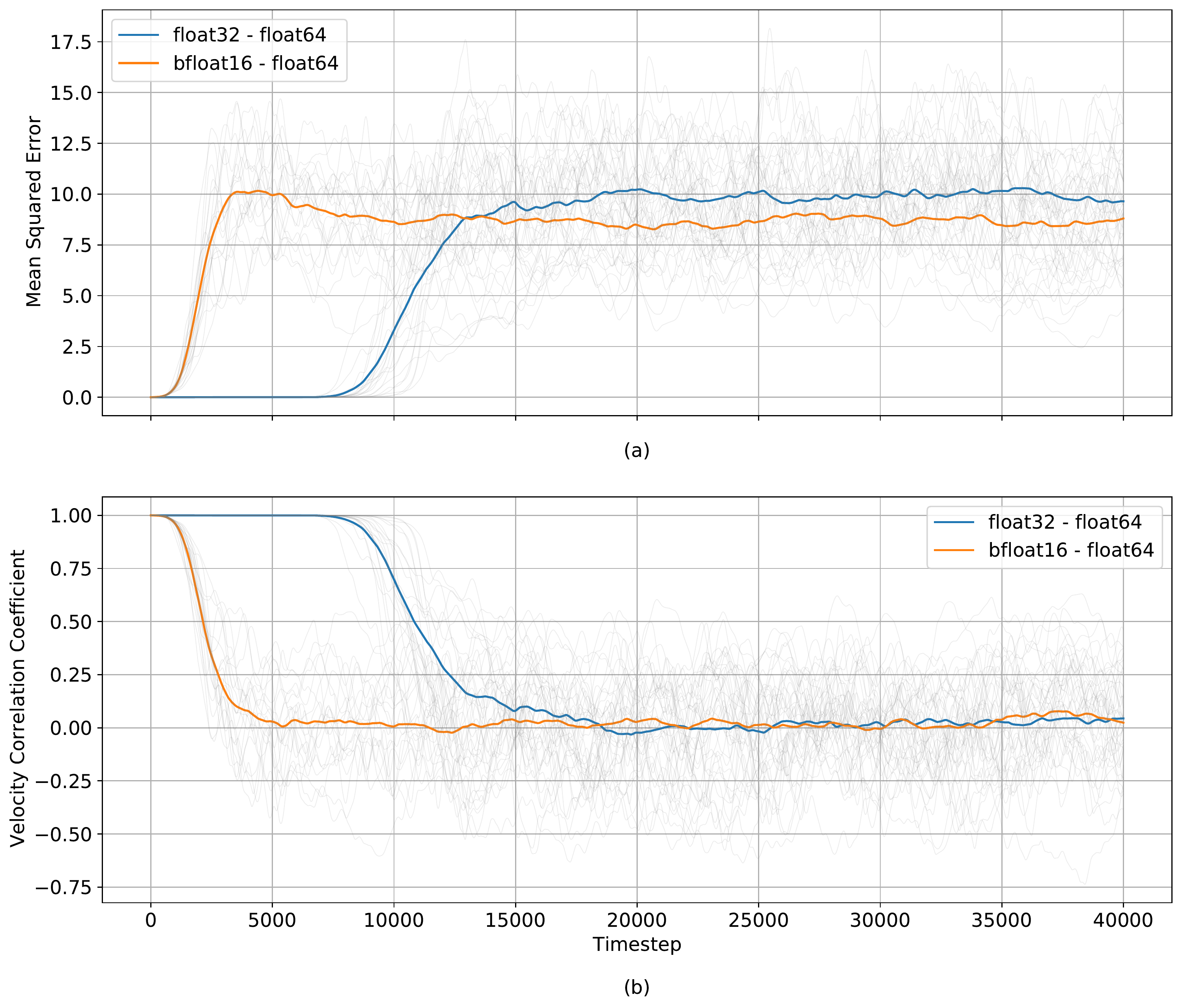}}
\caption{(a) Mean Squared Error accumulation in the velocity fields in a Kolmogorov forced turbulence test case run using half (bfloat16) and single (float32) precision arithmetic, compared to double precision (float64). (b) Correlation coefficient between the same velocity fields. Values are averaged over \begin{update}100\end{update} random initialisations.}
\label{fig:dataset_error}
\end{figure*}

We estimate the temporal evolution of the Mean Squared Error (MSE) of the velocity field between the reference 64-bit and the low-precision 32-bit/16-bit simulations for \begin{update}100\end{update} different random initializations of the flow field. In addition, we also estimate the statistical (Pearson) correlation between these velocity fields. Figures ~\ref{fig:dataset_error}(a) and (b) show the temporal evolution of the Mean Squared Error (MSE) and the correlation coefficient of the velocity fields, respectively. The variation of average MSE from these runs is also overlaid in the figure. As expected, the errors in the bfloat16 simulations set in sooner and at a faster rate than in the float32 simulations. Nevertheless, after a steep rise in the error, the MSE eventually saturates at some value and fluctuates about it. The saturation of MSE happens at the same time when the velocity fields obtained using different precision runs become completely uncorrelated, as seen in Figure ~\ref{fig:dataset_error}(b). Beyond this time, the pointwise metrics like MSE fail to capture the differences between the fields, which we explore further in Section~\ref{sec:results}. \begin{update} It is worth mentioning that MSE is more useful in the early part of the decorrelation process, where we can see that the bfloat16 error sets in significantly early and rises more steeply compared to float32. The average bfloat16 MSE saturates at a lower value than the average float32 MSE. The MSE saturation level for bfloat16 being marginally lower than that of float32 does not imply that bfloat16 is more accurate than float32. Once the fields have become completely decorrelated, i.e, the vortical structures are completely different, MSE is no longer a very useful metric to determine similarity. Similar observations have been made in the literature \cite{kohl2020learning}. \end{update}

\label{hardwarelimit1}
We measure the computational throughput of the solver on our hardware to be 6112 (steps/s) for float64, 10951 for float32 and 7891 for bfloat16. Although, in principle, bfloat16 should be faster than float32, it is observed to be slower than float32 because the hardware (V100 GPU) is not optimised for 16-bit operations \citep{gpubfloat16}. Due to the faster accumulation of error, we choose to use the 16-bit simulation for the rest of our experiments, where we use deep learning techniques to improve the accuracy of the predictions.

\begin{figure}[ptb]
\centerline{\includegraphics[width=0.5\textwidth]{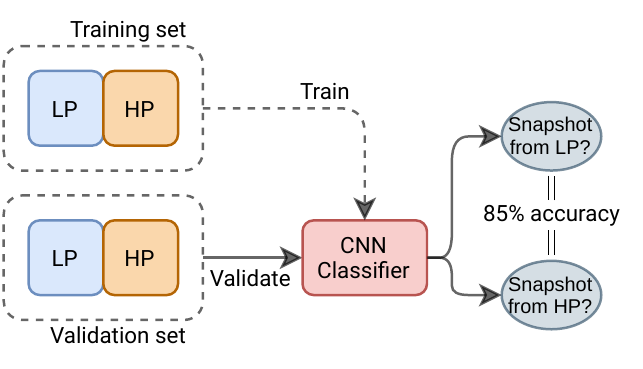}}
\caption{Preliminary experiment to test for the presence of generalisable structure in the flow produced by the low-precision solver. The LP and HP snapshots are generated using the same solver and flow parameters but with different numerical precision. The CNN classifies whether a snapshot came from the LP or HP solver with 85\% accuracy on a validation set.}
\label{fig:classifier}
\end{figure}

In order to use deep learning to learn a transformation from a low-precision flow into the corresponding high-precision flow, there must exist some generalisable structure to a flow snapshot determined by whether it was produced by the low- or high-precision solver. As a preliminary experiment to test for the presence of such a generalisable structure, we train a binary classifier on our data. In Fig.~\ref{fig:classifier}, the low-precision (LP) and high-precision (HP) snapshots are generated using the same solver and flow parameters but with different numerical precision (bfloat16 and float64, respectively). The CNN classifier is trained and subsequently applied on a held-out validation set. The trained classifier distinguishes low- and high-precision snapshots with 85\% accuracy, significantly greater than random. This result suggests that there is an inherent structure to the flow snapshots produced by the low-precision solver and that deep learning can model it.

%
%
%


%
%
%
\SetAlgorithmName{Pseudocode}{pseudocode}{List of Pseudocodes}
\begin{algorithm*}[ptb]
\caption{Generation of training data.}
\label{alg:datagen}
\KwIn{Time window $\Delta$, Correction frequency $N$, Dataset size $S$}

$T \gets \Delta / N$

\For{$\texttt{random\_seed} \gets 1$ : $S$}{

    \BlankLine
    \BlankLine
    
    \tcp{init transient discarded}
    $\textit{\texttt{init\_velocity}} \gets \texttt{initialization}(\textit{\texttt{random\_seed}})$
    
    \BlankLine
    \BlankLine
    
    \tcp{generate high-precision snapshots}
    $\textit{\texttt{solver\_precision}} \gets HIGH$
    
    $\textit{\texttt{snapshots}} \gets [\;]$
    
    $\textit{\texttt{velocity}} \gets \textit{\texttt{init\_velocity}}$
    
    \For{$\texttt{t\_outer} \gets 1$ : $N$}{
        \For{$\texttt{t\_inner} \gets 1$ : $T$}
            {$\textit{\texttt{velocity}} \gets \texttt{solver\_step}(\textit{\texttt{velocity}})$}
        Append $\textit{\texttt{velocity}}$ to $\textit{\texttt{snapshots}}$
    }
    
    \BlankLine
    \BlankLine
    
    \tcp{list containing $N$ snapshots}
    $\textit{\texttt{high\_prec\_snapshots}} \gets \textit{\texttt{snapshots}}$
    
    \BlankLine
    \BlankLine
    
    \tcp{generate low-precision snapshots, from same init}
    $\textit{\texttt{solver\_precision}} \gets LOW$
    
    Repeat steps \texttt{5} to \texttt{10} 
    
    $\textit{\texttt{low\_prec\_snapshots}} \gets \textit{\texttt{snapshots}}$
    
    \BlankLine
    \BlankLine
    
    Save to disk $\textit{\texttt{high\_prec\_snapshots}}$ and $\textit{\texttt{low\_prec\_snapshots}}$
}
\end{algorithm*}

\section{Methods to reduce error}
\label{sec:reduce}

In this section, we attempt to use deep learning to learn a function that can correct the error arising in a small duration of time between the low-precision and high-precision solvers. We use a neural network that interacts with the solver through differentiable programming \cite{um2020solver, kochkov2021machine}. 

For our experiments, we define $\Delta$ to be the total time duration (number of solver iterations) that the model sees in a single training sample. We define $N$ as the number of corrections the neural network makes to the low-precision simulation during the time duration $\Delta$. Therefore $T$, given by the relation $T = \Delta / N$, is the number of iterations the solver steps forward before passing the output to the neural network for a correction.

\subsection{Data generation}

The Pseudocode describes the process of generating data for training and validating the model. \begin{update}The initialization function takes a random seed as an argument and generates different realizations of synthetic turbulence having different vortical structures but the same spectral content. This is in contrast to generating different flow realizations by adding a random numerical noise to the flow field, which would in fact quickly dissipate. \end{update} The generated velocity field is filtered to have a maximum amplitude of 10 and a maximum wavenumber of 4. It then steps forward by 1500 timesteps and discards this initial transient. The returned velocity field is used to initialize both the high- and low-precision simulations by performing the appropriate typecasting. Starting from this initial condition, the velocity field is saved every $T$ iterations $N$ times, covering a total time duration of $\Delta = T \times N$. The simulation is repeated using both high- and low-precision numerics, which constitutes a single training sample of $N$ high and low-precision snapshots each. Note that the low-precision snapshots are only involved during the validation phase since, during training, they are generated on the fly by the hybrid solver.

\subsection{Model architecture}

\begin{figure*}[ptb]
\centerline{\includegraphics[width=\textwidth]{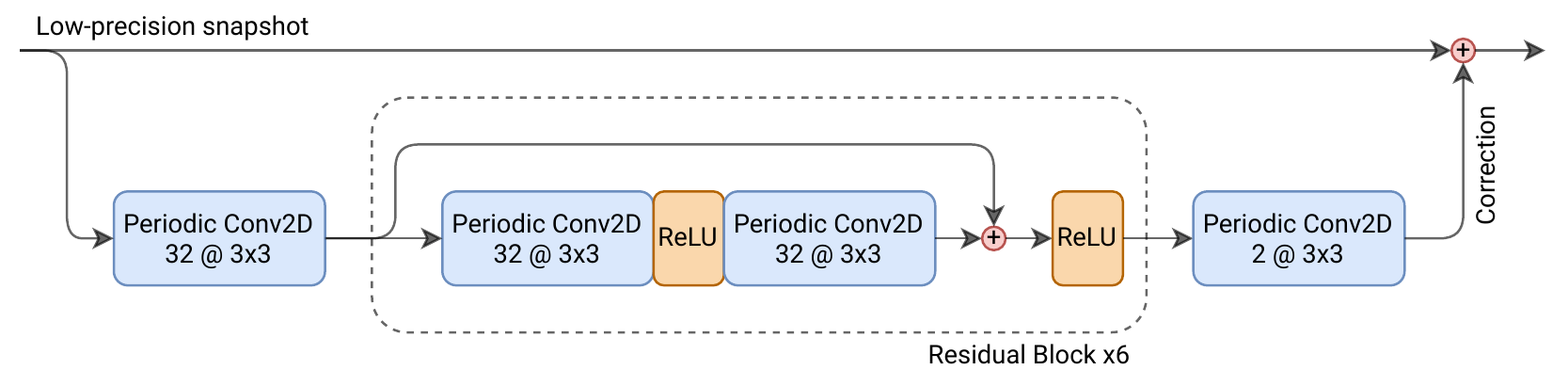}}
\caption{The Convolutional Neural Network architecture used to make corrections to the low-precision snapshot. C @ K $\times$ K denotes a convolutional layer with a K $\times$ K kernel and C output channels.}
\label{fig:cnn_arch}
\end{figure*}

We use a convolutional neural network (CNN) architecture consisting of 6 Residual Blocks (Fig.~\ref{fig:cnn_arch}). These blocks consist of 2D convolutional layers followed by ReLU activation functions and contain "skip connections", which significantly improve the gradient flow through the network \citep{resnet}. We use periodic convolutional layers instead of the standard convolutional layers used in deep learning tasks dealing with images. These layers mimic the periodic boundary conditions of the simulation domain by padding the edges of the input with values copied from the opposite edge. Furthermore, the input to a layer is padded such that the dimensions are maintained in the output. The network learns a residual correction to the input low-precision snapshot, which is added to the input to produce the corrected snapshot. 

We employ this neural network to periodically apply a correction to the output of the low-precision solver before passing it back to the solver for the next iteration. Both the solver and the neural network are implemented in JAX, which supports reverse-mode automatic differentiation. Therefore, the gradient of the loss function w.r.t the network weights can be backpropagated through the solver operations, allowing the neural network and solver to work in tandem. The unrolling of this hybrid solver and the subsequent gradient calculation are similar to the procedure for a recurrent neural network (RNN). 

\begin{figure*}[ptb]
\centerline{\includegraphics[width=\textwidth]{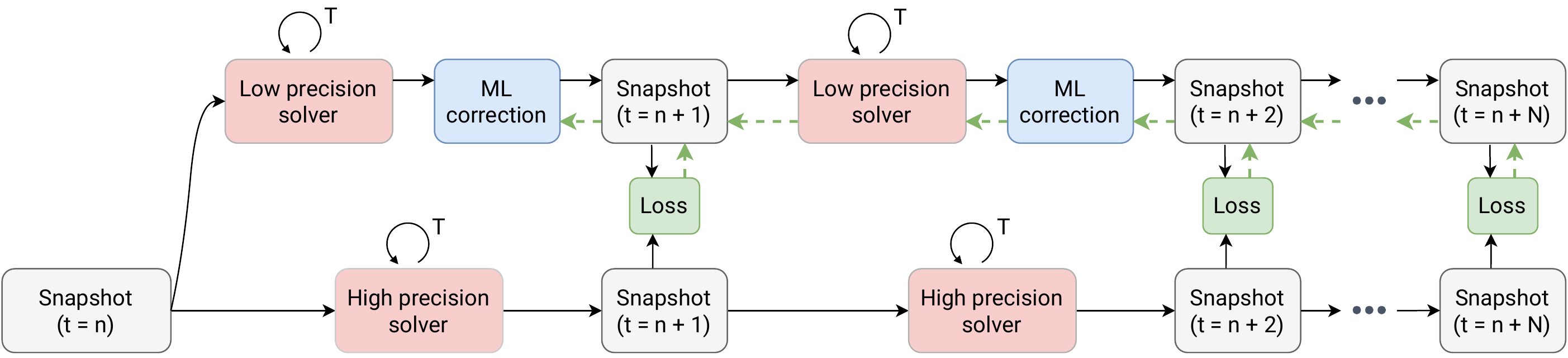}}
\caption{Schematic showing how the hybrid model is trained. The solid arrows indicate the forward simulation process. The top row shows the hybrid solver (to be trained), and the bottom row shows the reference high-precision solver. The dashed arrows indicate the backpropagation of the gradient through the unrolled differentiable solver and neural network.}
\label{fig:model_schematic}
\end{figure*}

Figure~\ref{fig:model_schematic} shows a schematic of how the hybrid model is trained. The corrected outputs of the neural network at each iteration are saved and stacked together. The loss is calculated as the Mean Squared Error (MSE) between the corrected and corresponding high-precision snapshots. The gradient of this loss is backpropagated as shown, and the network weights are updated to minimise the loss. 

We study the effect of the hyperparameters $T$, $N$ and $\Delta$ in detail in Section~\ref{sec:results}. For the model and optimisation hyperparameters, we stick to simple default values as \citet{kochkov2021machine} report that their model performance does not have a consistent dependence on these parameters. We use the Adam optimisation algorithm \citep{adam} with a learning rate of 1e-4 and batch size of 2, and each model takes around 24 hours to train on our hardware.

\subsection{Advantages of the hybrid solver}
\label{sec:model}
\begin{figure}[ptb]
\centerline{\includegraphics[width=0.4\textwidth]{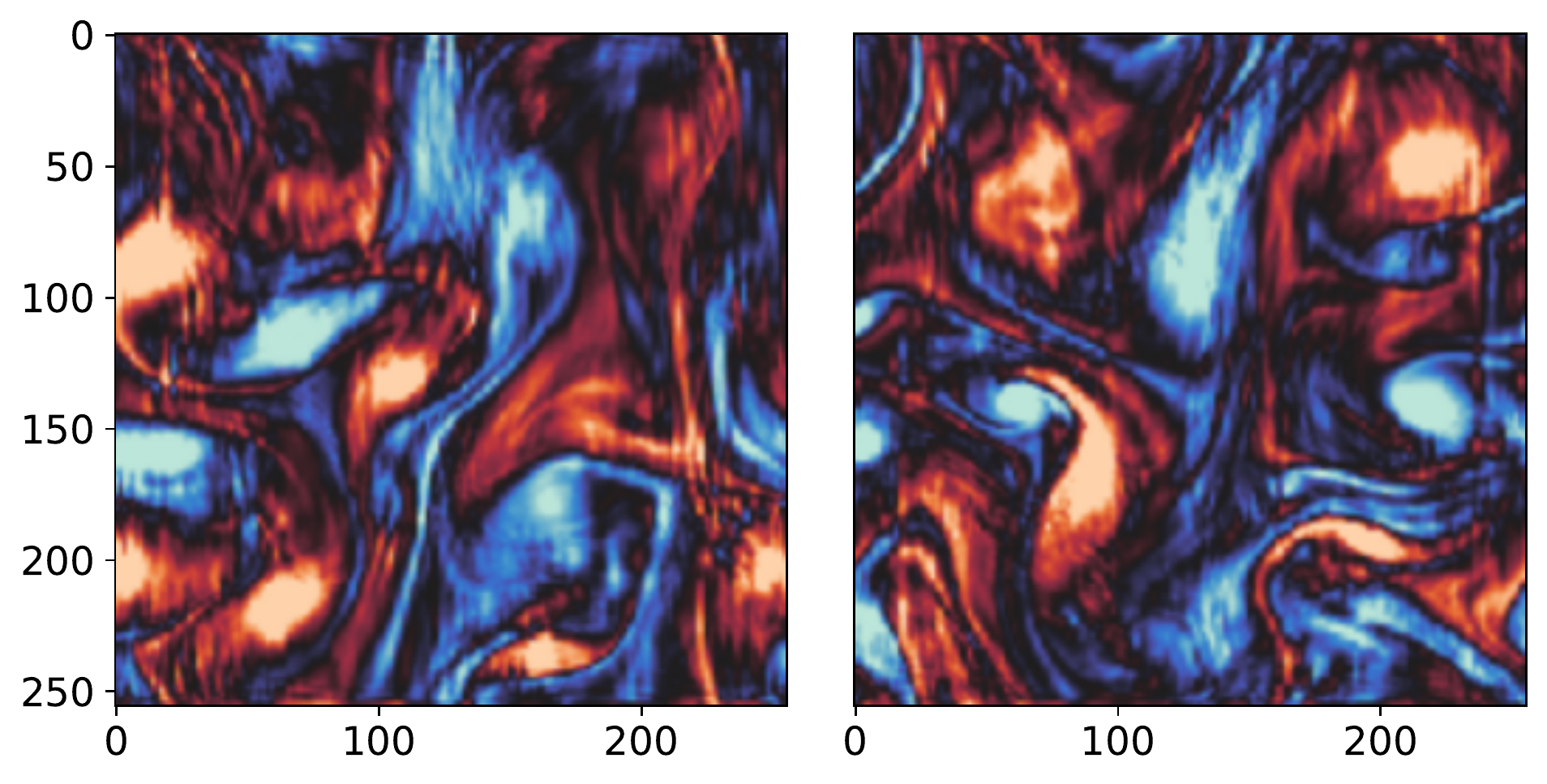}}
\caption{Typical visualisation of the vorticity fields obtained from a pure ML model trained on a pre-computed dataset in a supervised manner. Two example snapshots are shown here.}
\label{fig:post_proc}
\end{figure}

The advantages of this hybrid solver approach over a pure ML supervised learning approach are as follows. First, physics constraints such as conservation of mass and momentum are more easily achieved since the actual time stepping is performed by the physics solver, not by the neural network. Second, when a correction is computed, the input to the neural network is a flow field that resulted from a correction by the same neural network at an earlier timestep. Furthermore, the current correction will affect inputs to the neural network correction function at later timesteps. This allows the neural network to account for the future performance of a correction during the learning process. In a standard supervised learning approach, by contrast, the inputs to the neural network are pre-computed prior to training and are not affected by past corrections in any way \citep{um2020solver, kochkov2021machine, hybridreactive}. Figure~\ref{fig:post_proc} visualises two example timesteps produced by such a model. The output is highly distorted and physically inaccurate. Therefore we do not pursue further experiments using the standard supervised learning approach. In the subsequent section, we demonstrate the improvements produced by the hybrid solver over this method.

%
%
%

\section{Experiments and Results}
\label{sec:results}

We experiment with training the model on data generated using $\Delta = 150$ and $\Delta = 300$. In each case, we experiment with four values of $N$ (the number of corrections made during $\Delta$). For each set of values $\{\Delta, N\}$, we generate a training, validation and held-out test set with dataset size $S$ = 200, 10 and 2, respectively. The validation and held-out test sets are generated as in the Pseudocode, whereas in the case of training data, the low-precision snapshots are generated on-the-fly during training.

While the models see windows of 150 and 300 timesteps during training, we test the error reduction and stability of the model outputs over a longer time duration. We generate a \textit{long-run} test set consisting of simulation sequences of several thousand timesteps, allowing us to test the ability of the trained models to produce corrected sequences far longer than what they saw during training. 

\begin{table*}[ptb]
\centering
\small
\caption{Results of models trained over sequences of $\Delta=150$ timesteps, averaged over multiple test set samples.}
\label{tab:results150}
\begin{tabular}{@{}c c p{0.4\linewidth} p{0.1\linewidth} p{0.15\linewidth}@{}}
\toprule
\multicolumn{5}{c}{$\boldsymbol{\Delta = 150}$} \\ \midrule
$T$ & $N$ & \centering Rel. improvement in MSE over $\Delta$ & \centering $t_{90\%}$ (steps) & {\centering Throughput (steps/s)}  \\ \toprule\midrule
75      & 2      & \centering 12.3\%   & \centering 315     &      2900\\ 
10      & 15     & \centering 38.8\%     & \centering 2158     &      1270\\ 
5       & 30     & \centering 43.6\%     & \centering 1353     &      675\\ 
3       & 50     & \centering 47.6\%     & \centering 1894     &      423\\ \bottomrule
\end{tabular}
\end{table*}
\begin{table*}[ptb]
\centering
\small
\caption{Results of models trained over sequences of $\Delta=300$ timesteps, averaged over multiple test set samples.}
\label{tab:results300}
\begin{tabular}{@{}c c p{0.4\linewidth} p{0.1\linewidth} p{0.15\linewidth}@{}}
\toprule
\multicolumn{5}{c}{$\boldsymbol{\Delta = 300}$} \\ \midrule
$T$ & $N$ & \centering Rel. improvement in MSE over $\Delta$ & \centering $t_{90\%}$ (steps) & {\centering Throughput (steps/s)}  \\ \toprule\midrule
150      & 2      & \centering 13.4\%   & \centering -     &      3150\\ 
20      & 15     & \centering 41.9\%     & \centering 1608     &      1720\\ 
10       & 30     & \centering 45.7\%     & \centering 1324     &      1270\\
6       & 50     & \centering 29.3\%    & \centering 1658     &      785\\ \bottomrule
\end{tabular}
\end{table*}

The metrics we use to quantify the performance of the models broadly fall into two categories: a) aggregate metrics such as kinetic energy spectra and mean divergence, which deal with statistical accuracy of the flow, and b) error metrics such as Mean Squared Error (MSE) and correlation coefficient which deal with pointwise accuracy of the flow, i.e., the exact evolution of eddy locations over time. 

Tables~\ref{tab:results150} and \ref{tab:results300} show the relative improvement in MSE during a time duration $\Delta$, which is calculated on the held-out test set for each $\{\Delta, N\}$. $t_{90\%}$ measures the number of timesteps for which the MSE between the corrected sequence and the high-precision sequence is $\leq90\%$ of the MSE between the low-precision sequence and the high-precision sequence.

\begin{figure*}[ptb]
\centerline{\includegraphics[width=\textwidth]{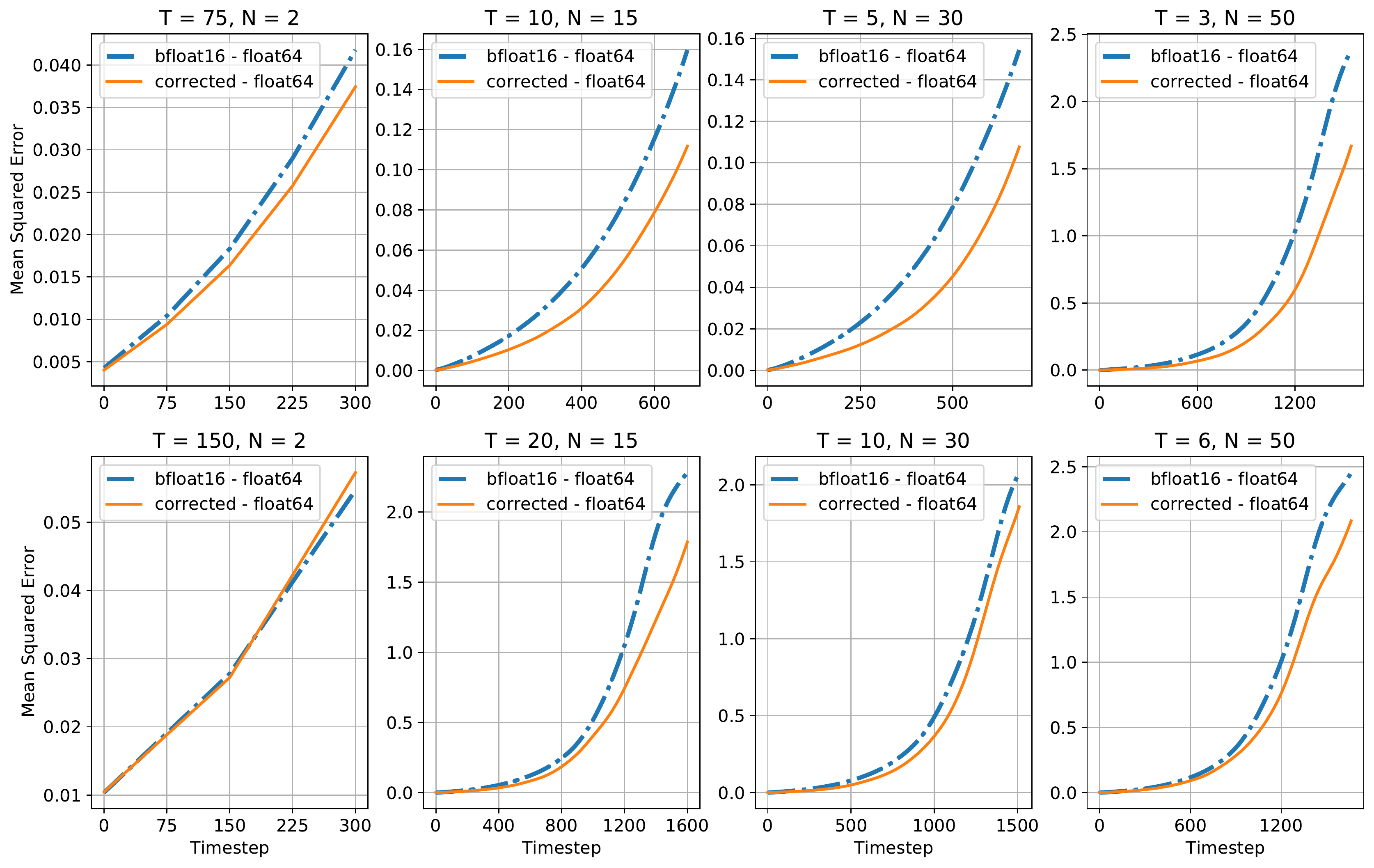}}
\caption{Reduction in Mean Squared Error achieved by the trained models when unrolled over long time durations, averaged over 5 sequences from the test set. The top and bottom rows show $\Delta=150$ and $\Delta=300$ experiments respectively (recall that $T \times N = \Delta$).}
\label{fig:pred_evolution}
\end{figure*}

We observe that the models making more frequent corrections ($N = 30$ and $N = 50$) perform better in terms of relative improvement in MSE over $\Delta$, compared to the models making infrequent corrections ($N=2)$.  In terms of $t_{90\%}$, the $N = 2$ models fail to keep the error under control for very long. The $N=15$, $N=30$ and $N=50$ models, on the other hand, have $t_{90\%}$ significantly greater than the time duration which they saw during training. The $T=6$, $N=50$ model shows an anomalous value of rel. improvement in MSE, despite which it has a better $t_{90\%}$. Figure~\ref{fig:pred_evolution} shows the evolution of the MSE for the low-precision and corrected sequences on the \textit{long-run} test set. Except for the $T=150, N=2$ model, all the models delay the error accumulation by some amount. Looking beyond the timesteps shown in Fig.~\ref{fig:pred_evolution}, however, all the flows eventually become uncorrelated with their target high-precision flows.

\label{hardwarelimit2}
We measure the effective computational throughput of each of our learned solvers on our workstation using a single NVIDIA V100 GPU. For a given $\Delta$, the more frequently the neural network is invoked to make a correction, the more computational overhead is introduced. This presents a trade-off, which can be seen in Tables~\ref{tab:results150} and \ref{tab:results300}. Making more frequent corrections results in lower error but incurs greater computational cost resulting in lower throughput. The throughput can be further improved by training and evaluating the neural network using bfloat16 arithmetic, which has been shown to preserve their learning capability \citep{kalamkar2019study}. However, due to the lack of support for bfloat16 on the Tensor Cores on our available hardware, \citep{gpubfloat16} this could not be done in the present study. For this same reason, a direct comparison of throughput with the base 16-bit solver is not meaningful.

\begin{figure*}[ptb]
\centerline{\includegraphics[width=\textwidth]{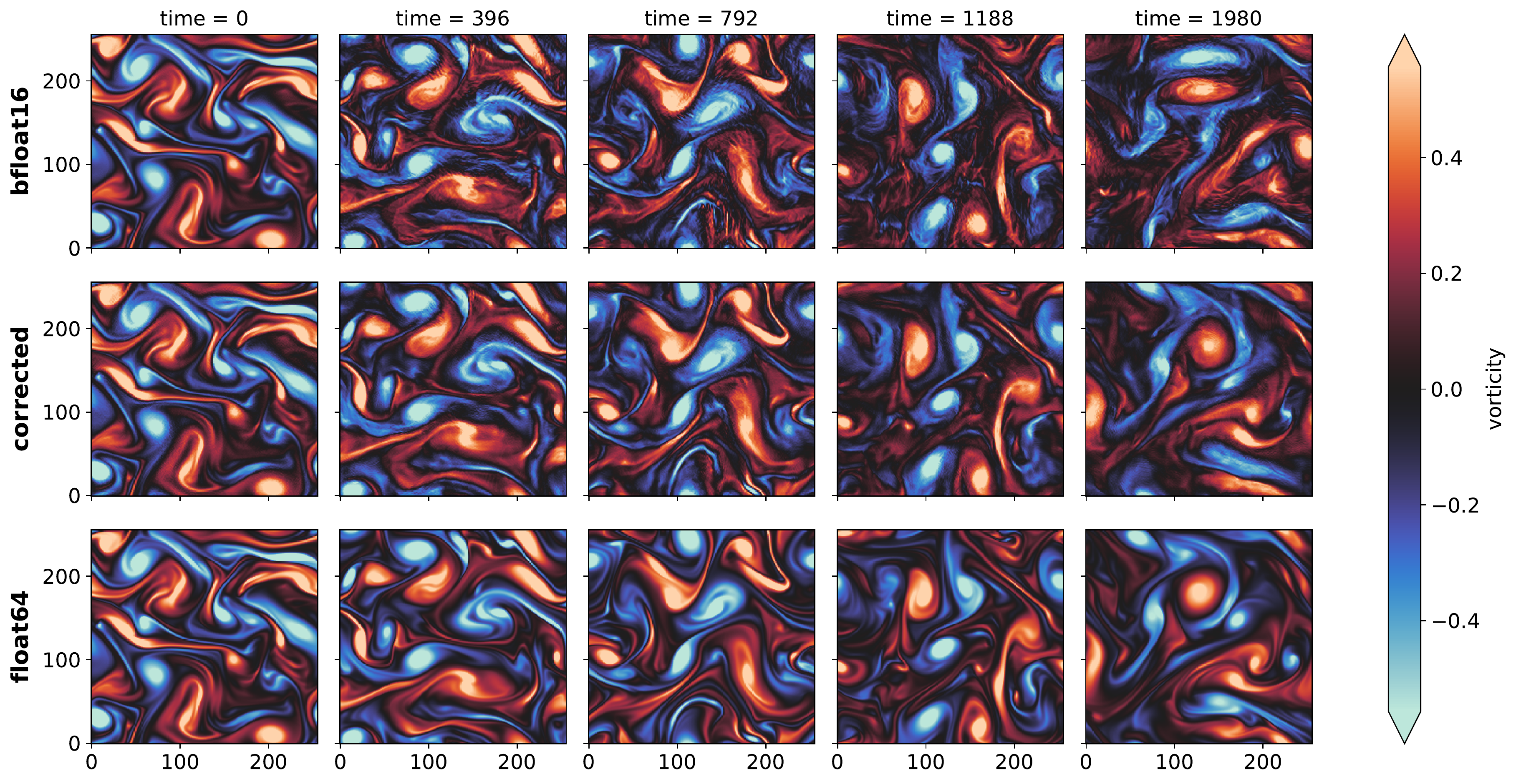}}
\caption{Typical comparison of the vorticity fields obtained from the present model ($\Delta = 150, N = 50$ condition), low-precision and high-precision simulations, as a function of time. Beyond the timestep shown here, the corrected and reference flows start to gradually become uncorrelated.}
\label{fig:pred_example}
\end{figure*}

Figure~\ref{fig:pred_example} visualizes the vorticity fields obtained from the experiment with $\Delta = 150, N = 50$. The low-precision flow appears grainy and distorted, whereas the corrected flow regains the lost visual quality. Furthermore, the locations and shapes of the vortices in the final snapshot match more closely with the corresponding high-precision snapshot, which indicates that the corrective model delays the time when the low- and high-precision flows become uncorrelated (see snapshots at $t=1980$).

\begin{figure*}[ptb]
\centerline{\includegraphics[width=\textwidth]{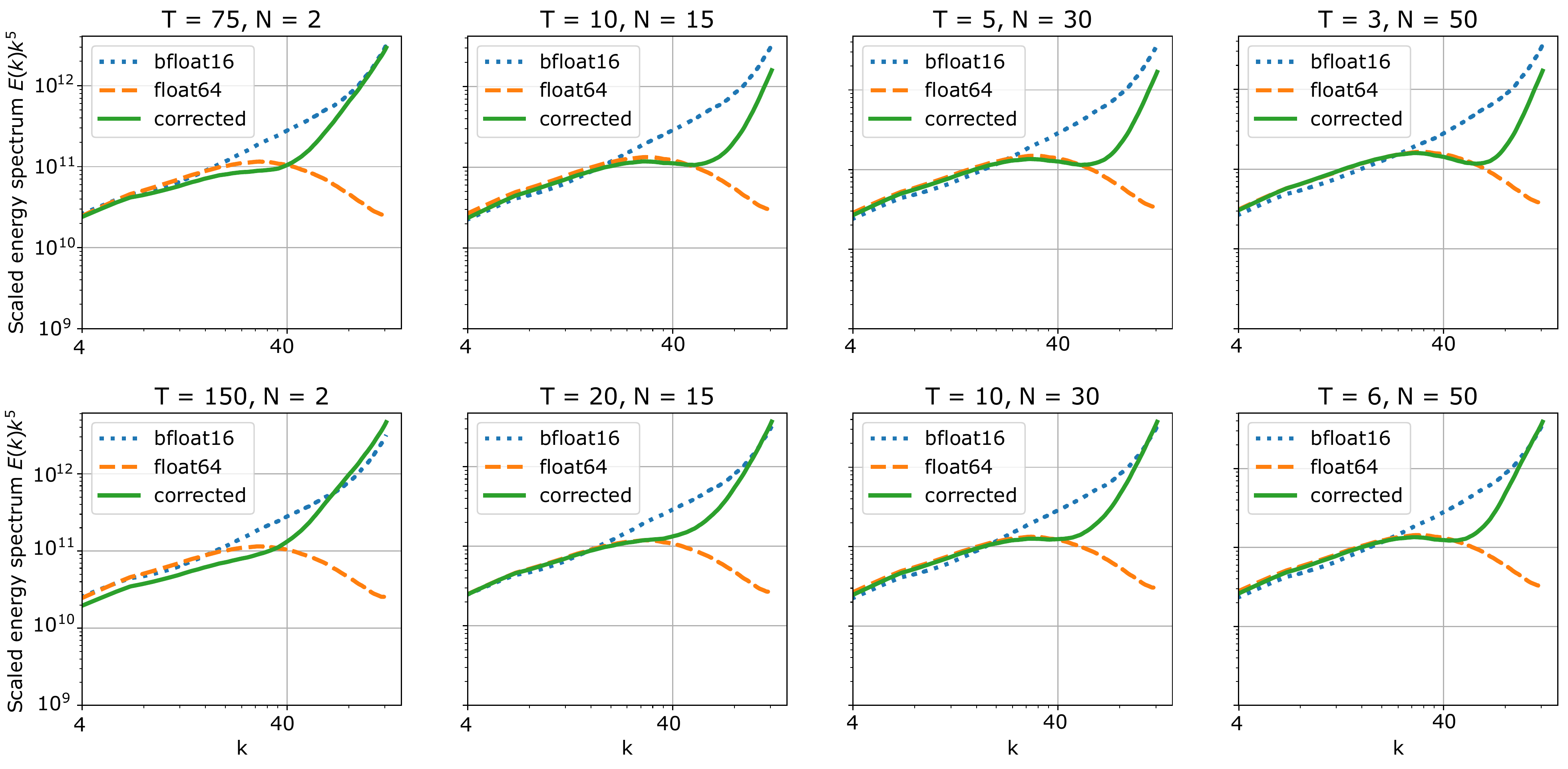}}
\caption{Comparison of the kinetic energy spectra scaled by $k^5$ of the low-precision, high-precision and corrected sequences, averaged over 5 sequences of 1000 snapshots each from the test set. The top and bottom rows show $\Delta=150$ and $\Delta=300$ experiments respectively.}
\label{fig:spectrum_comparison}
\end{figure*}

Mean Squared Error, being a pointwise comparison metric, only considers local distances and may fail to capture structural similarity and physical characteristics of numerical simulations \citep{kohl2020learning}. Therefore it is not a sufficient metric on its own to capture some of the differences seen in Fig.~\ref{fig:pred_example}, which can be better characterised by inspecting flow properties such as the kinetic energy spectrum. Therefore, we compare the kinetic energy spectra $E(k) = \frac{1}{2}\vert u(k) \vert^2$ of the low-precision, corrected and high-precision sequences in each experiment. Figure~\ref{fig:spectrum_comparison} shows the average of the scaled kinetic energy spectrum $E(k) \cdot k^5$ over 1000 snapshots from the \textit{long-run} test set. These 1000 snapshots are taken between 3000 and 150000 timesteps. From Fig.~\ref{fig:dataset_error}, it is evident that the corrected and reference flows are uncorrelated at these timesteps. Generally, both the low-precision and corrected sequences yield spectra which match closely with the high-precision spectra in the low to mid frequency range ($k=4\text{--}20$). However, we see that the graininess caused by low numerical precision results in surplus energy accumulation at high frequencies (larger $k$). The models trained with $N=2$ fail to produce accurate spectra compared to the high-precision simulation at all frequency ranges. The learned solvers with more frequent corrections ($N = 30$ and $N = 50$) produce significantly improved spectra. They reduce the surplus high-frequency energy accumulation, except at the highest frequencies, where all the models deviate from the high-precision spectra.

We compute the divergence of the velocity fields in the high-precision, low-precision and corrected sequences using finite difference derivatives and inspect its mean value over the domain. We observe that all three flows have a mean divergence of order 1e-11, which shows that the correction made by the neural network preserves the net-zero divergence property of the field.

\begin{figure*}[ptb]
\centerline{\includegraphics[width=0.9\textwidth]{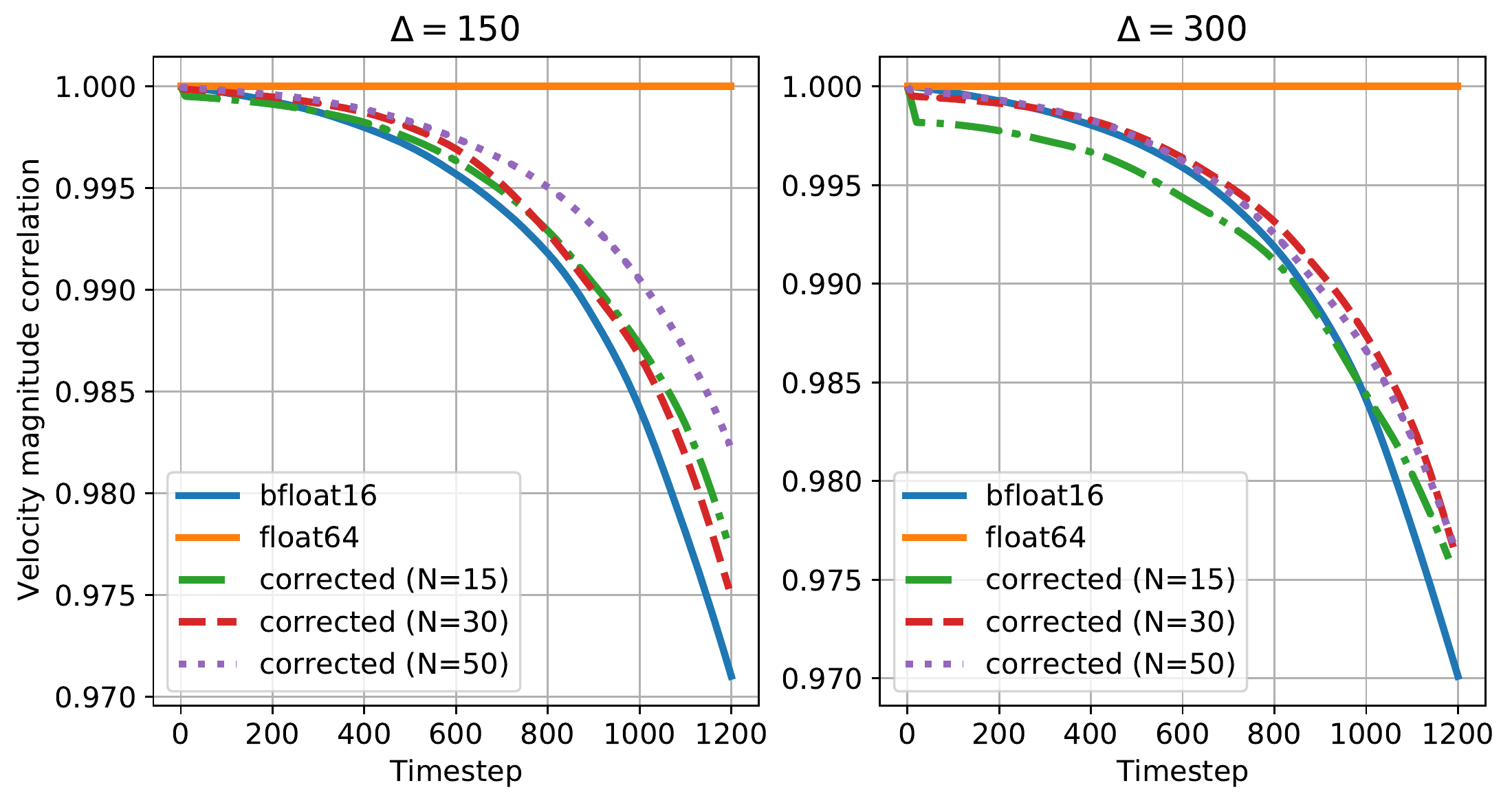}}
\caption{The velocity magnitude correlation between the corrected and the high-precision flows as a function of time, averaged over 5 sequences from the test set.}
\label{fig:velocity_correlation}
\end{figure*}

Figure~\ref{fig:velocity_correlation} shows the statistical correlation of the velocity magnitude between the corrected and the high-precision flows as a function of time. In this case, as well, the models making more frequent corrections perform better. With the exception of the $\Delta = 300, N = 15$ and all the $N = 2$ models, the corrections made to the low-precision flow delay its decorrelation with the reference flow. Compared to the coarse-grid simulation of \citet{kochkov2021machine}, the rate of decorrelation of the low-precision simulation is significantly slower. Consequently, our models achieve a smaller relative delay in decorrelation than their models.

\begin{figure*}[ptb]
\centerline{\includegraphics[width=0.5\textwidth]{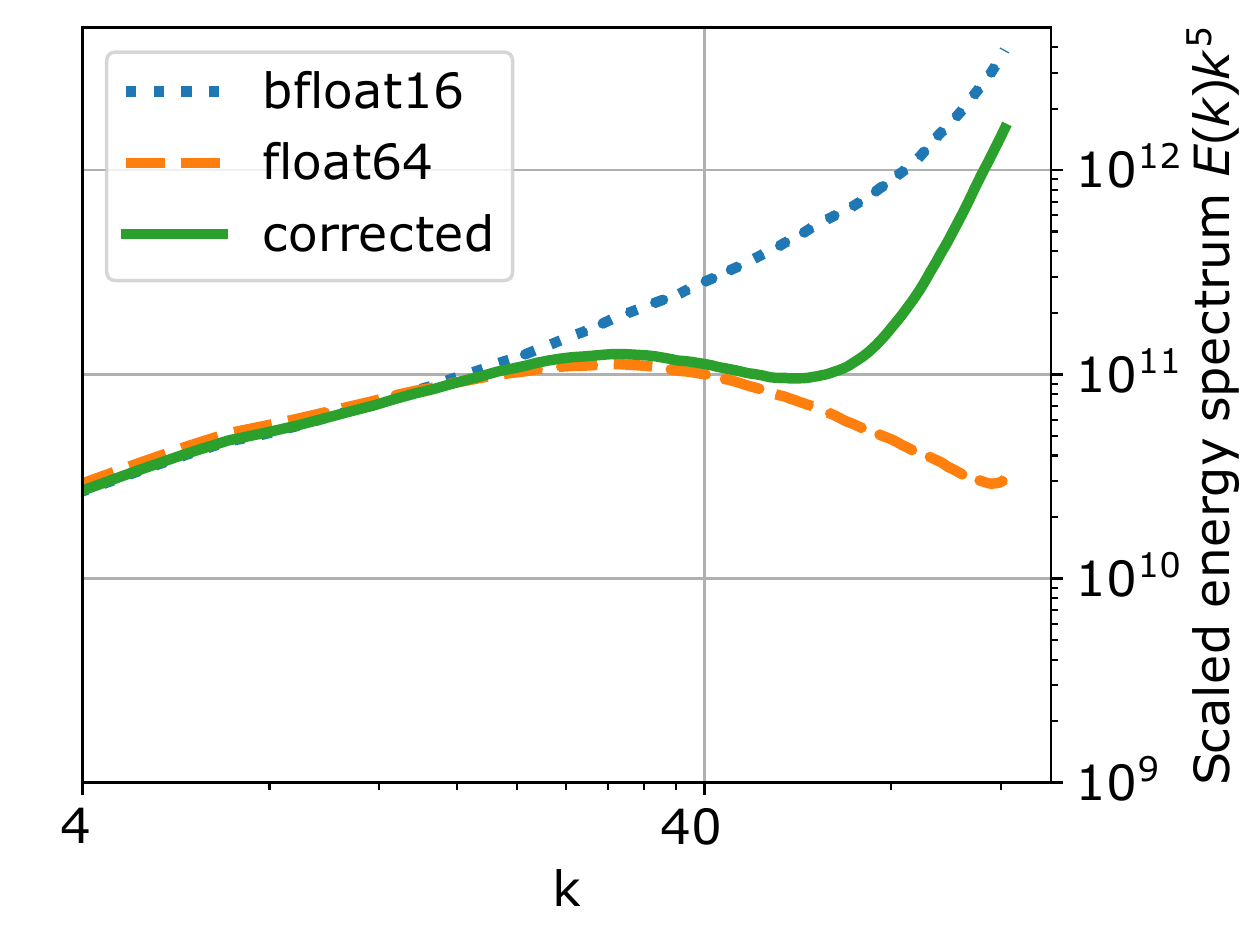}}
\caption{Comparison of the kinetic energy spectra scaled by $k^5$ obtained by applying the $T = 3, N = 50$ trained neural network (without further training) to a simulation performed at a 10 times higher Reynolds number.}
\label{fig:Re40000_pred_example}
\end{figure*}

As a test of generalization, we take the neural network obtained using the best-performing set of hyperparameters ($T=3, N=50$) and use it to generate corrections to a simulation performed at a 10 times higher Reynolds number (Re=40000 instead of 4000). Figure \ref{fig:Re40000_pred_example} compares the KE spectra with and without NN correction. With no further training, the corrections made by the neural network to the Re=40000 low-precision solver's output achieved similar improvements in all the metrics (Rel. improvement in MSE over $\Delta$, $t_{90\%}$, KE spectrum, etc.). This suggests that in addition to identifying trends in the model's performance, we have also identified a set of hyperparameters that can be used as a useful starting point when applying our method to different flows. 

There are broadly two kinds of metrics used in this study to evaluate the performance of NN corrections: a) aggregate metrics such as KE spectra and mean divergence which deal with the statistical accuracy of the flow, and b) error metrics such as Mean Squared Error and correlation coefficient which deal with pointwise accuracy of the flow, i.e, the exact evolution of eddy locations over time. \begin{update} From Figure \ref{fig:spectrum_comparison}, the role of the proposed NN correction appears to be identical to that of a filter that damps energy content over a range of frequencies. An experiment has been performed (refer to \ref{sec:appendix}) in these lines where we have compared the spectra and vortical structures replacing the NN correction with a spectral filter. The physics-informed NN correction function learned from data is observed to be superior when compared to a simple filtering operation. Further investigations can explore the efficacy of a combination of NN correction (to
control mid-high frequencies) with a standard filtering operation (to control
high-frequency spectra), considering the computational overheads. \end{update}

\section{Limitations}
\begin{update}
While our method achieves improvements in error metrics and physical quantities, the hybrid solver does not beat the reference high-precision solver in computational throughput. This is due to limitations in hardware available to carry out the study, as mentioned in Sections~\ref{hardwarelimit1} and \ref{hardwarelimit2}. The computational efficiency can be further optimized by running on hardware designed for low-precision computation, training the neural network itself using bfloat16 \cite{kalamkar2019study}, applying convolutions in the Fourier domain \cite{li2020fourier}, and applying model quantization techniques \cite{quantization}. Improving the throughput of the high-precision solver is an important direction for future work.
\end{update}

The motivation for this method stems from related works by \citet{kochkov2021machine} and \citet{um2020solver}. In both these studies and in the current study, snapshots from a Navier-Stokes solver are used as the targets during training, which means that the neural network is implicitly trained to make corrections that satisfy the governing equations. A limitation to this strategy is that the corrections themselves need not explicitly satisfy the governing equations of the flow. Instead, by using snapshots from a real solver as the targets during training, the neural network is implicitly trained to predict a flow field that satisfies the governing equations. A direction for future work that addresses this limitation is to combine our method with the method proposed by \citet{pinn}, where the network's output is explicitly regularised to satisfy the governing equations by including an additional physics-based term in the loss function.

\section{Conclusion}
\label{sec:conclusion}
In this work, we explored the error accumulation due to low precision in a Kolmogorov forced turbulence test case and used deep learning together with a differentiable solver to improve the simulation. The learned ML-CFD hybrid reduces the Mean Squared Error accumulation over time durations longer than it saw during training, improves the kinetic energy spectra and preserves the mean divergence of the flow. We show that making more frequent corrections to the simulation achieves a greater reduction in error, although it comes with an increased computational cost. The neural network's corrections improve the visual fidelity of the low-precision flow, making it look less grainy and distorted.
Furthermore, the correction is dependent on only 2 parameters (the frequency of corrections and total time window seen during training), apart from the usual parameters involved in neural network training. We quantified the dependence of the model's performance on these parameters and identified trends which will help implement the method in reality. The GPU memory savings obtained from the reduced numerical precision can enable simulations with more mesh elements than would otherwise be possible. Furthermore, since the hybrid solver is fully differentiable, it is suitable for use in control and inverse problems. While established industrial solvers written in FORTRAN or C are not directly compatible with our deep learning method implemented in JAX, new high-order accurate differentiable solvers are under development using the JAX framework \citep{jaxsolver, jaxhpc}, which solidifies the long-term viability of our method in real-world uses. We discussed the limitations of the method and identified opportunities to extend and improve it in future work. \begin{update} Although the current study focuses on non-reacting turbulent flows, the methods and the neural network framework discussed here can be readily extended to reacting flows. \end{update}

\section*{Acknowledgments}

N.R.V acknowledges the Department of Science and Technology-Science and Engineering Research Board (DST-SERB) for funding the project. The support and the computing resources provided by PARAM Sanganak under the National Supercomputing Mission, Government of India, and P G Senapathy Centre, IIT Madras are gratefully acknowledged. Y.M. acknowledges the support of Japan Science and Technology Agency, PRESTO, Grant Number JPMJPR21OB.


\begin{update}

\appendix

\section{Comparison with a simple spectral filter}
\label{sec:appendix}

We performed an experiment where we replaced the NN correction with a spectral filter. The filter was applied every T=3 solver iterations, the same as the best-performing NN correction in our experiments. Figures~\ref{fig:filter_vorticity} and \ref{fig:filter_spectra} show the visualization of the vorticity and the KE spectra resulting from this experiment.

The spectral filter \cite{spectralfilter} is formulated as:

\[\alpha_f \overline{\phi}_{i-1} + \overline{\phi}_i + \alpha_f \overline{\phi}_{i+1} = \frac{(\frac{1}{2} + \alpha_f)}{2} (2 u_i + u_{i-1} + u_{i+1})\]

where $\phi$ is the filtered velocity field and $u$ is the unfiltered velocity field. The strength of the filter is controlled by the tunable parameter $\alpha_f$ in the range (-0.5, 0.5), with $\alpha_f$ closer to 0.5 resulting in minimal filtering. The results for different values of $\alpha_f$ are shown in \ref{fig:filter_spectra} and Figures~\ref{fig:filter_vorticity}. From the spectra, it is evident that higher values of filter coefficient ($\alpha_f = 0.49$) damp energy at higher wavenumbers, while a lower filter coefficient ($\alpha_f = 0$) damps the energy content over a wide range of wavenumbers. The physics-informed NN correction function learned from data is superior when compared to a simple filtering operation. The visualizations of the vorticity field in Figure ~\ref{fig:filter_vorticity} further confirm this observation where the strength of the vortices also decreased with increasing $\alpha_f$ affecting the entire KE spectra. It is worth mentioning that all the sub-grid filters known to the authors are ‘grid based’ (which use the filter width based on the grid size) while the current study focus accumulation of the ‘precision-based’ errors. Further investigations can be carried out along these lines to explore the possibility of developing `precision-based' filters. The efficacy of a combination of NN correction (to control mid-high frequencies) with a standard filtering operation (to control high-frequency spectra), considering the computational overheads, can also be verified.

\begin{figure*}[ptb]
\centerline{\includegraphics[width=0.5\textwidth]{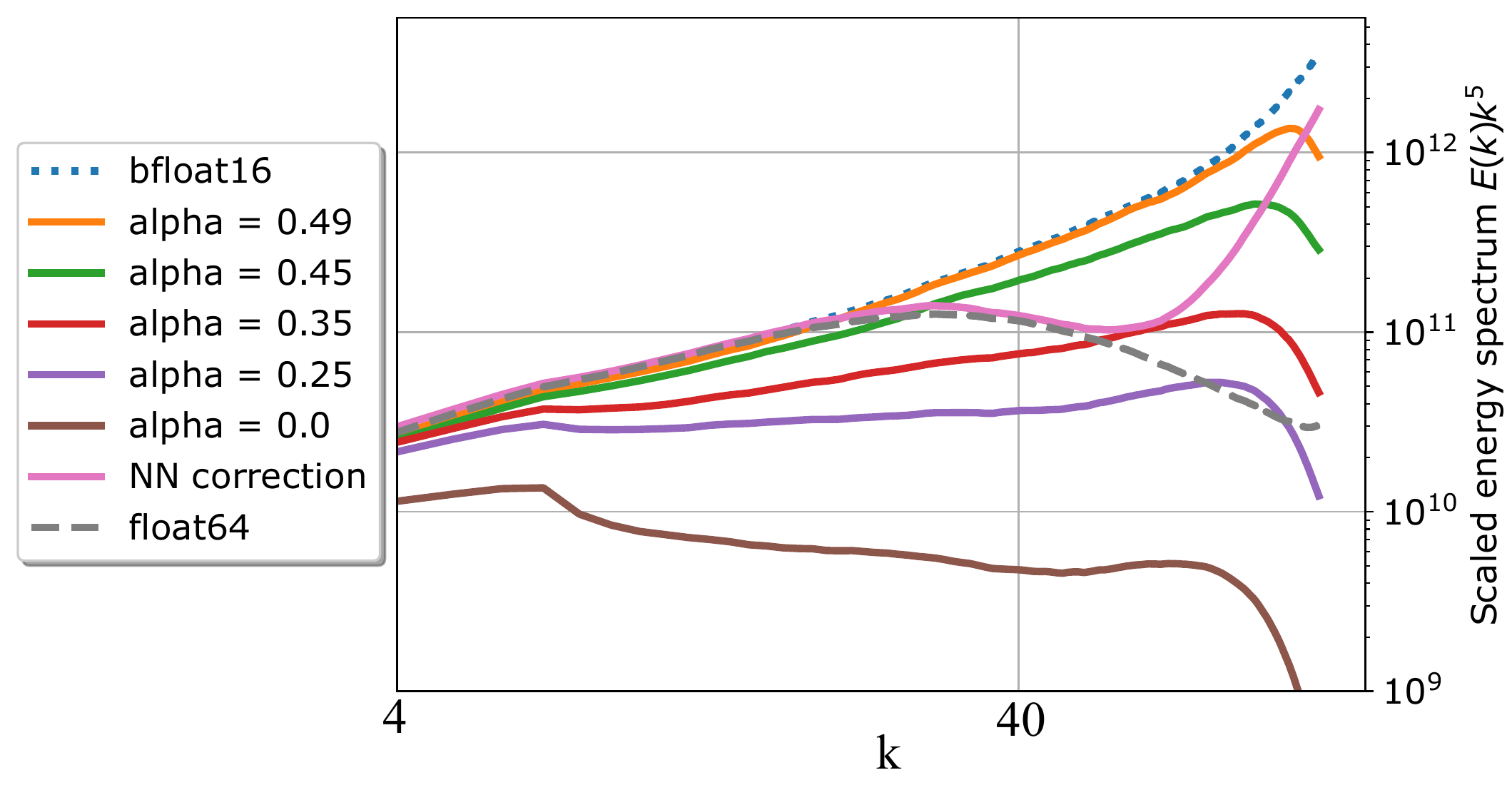}}
\caption{Comparison of the kinetic energy spectra scaled by $k^5$ obtained by replacing the neural network correction with a spectral filter with different values of $\alpha_f$.}
\label{fig:filter_spectra}
\end{figure*}

\begin{figure*}[ptb]
\centerline{\includegraphics[width=\textwidth]{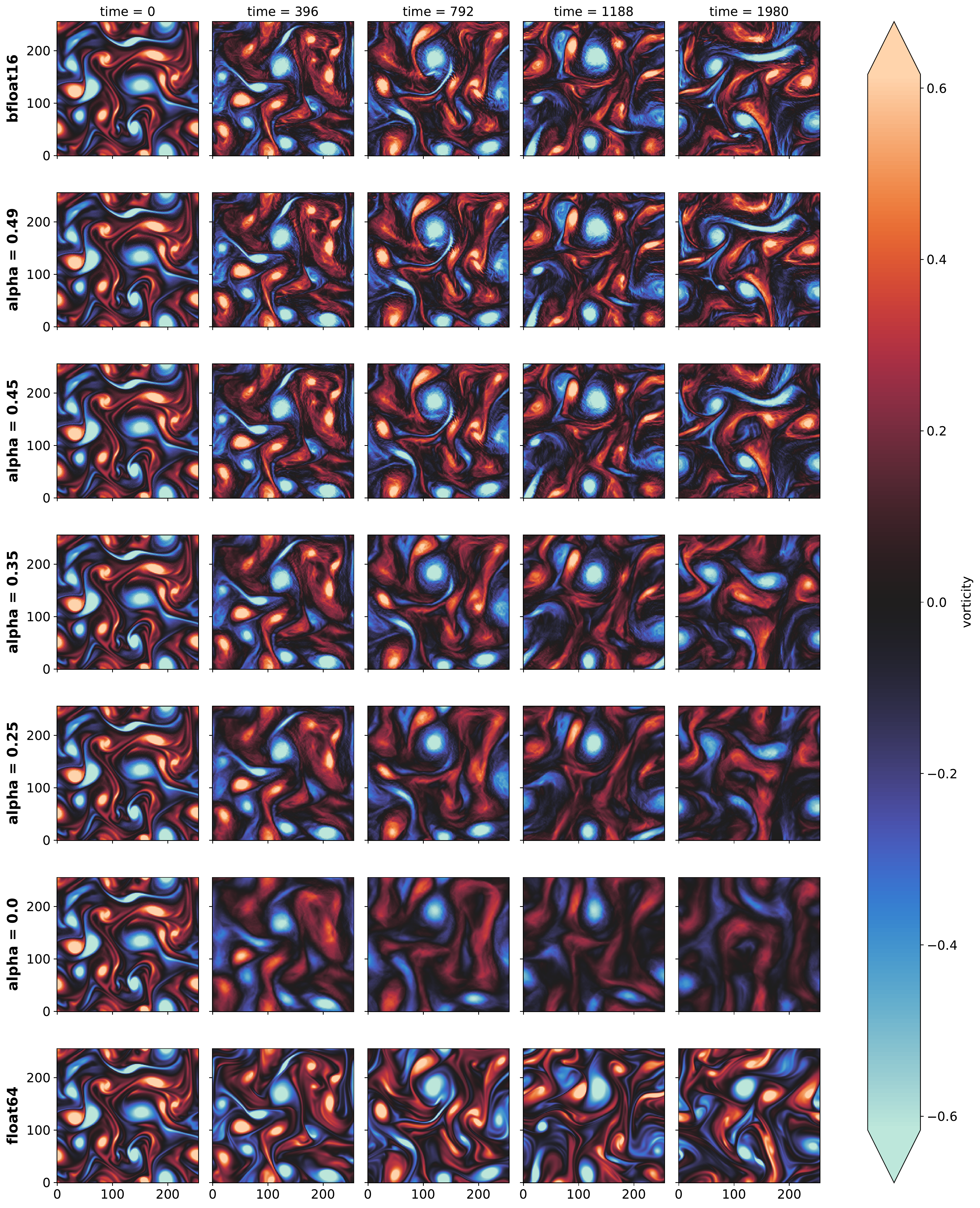}}
\caption{Visualization of the voriticity fields obtained by replacing the neural network correction with a spectral filter with different values of $\alpha_f$.}
\label{fig:filter_vorticity}
\end{figure*}

\end{update}

\bibliographystyle{unsrtnat} 
\bibliography{cas-refs}
 
\end{document}